\documentclass[conference]{IEEEtran}
\IEEEoverridecommandlockouts

\usepackage{cite}
\usepackage{amsmath,amssymb,amsfonts}
\usepackage{algorithmic}
\usepackage{graphicx}
\usepackage{textcomp}
\usepackage{xcolor}
\usepackage{multicol}
\usepackage{multirow}
\usepackage{booktabs}
\usepackage{url}
\usepackage{balance}

\def\BibTeX{{\rm B\kern-.05em{\sc i\kern-.025em b}\kern-.08em
    T\kern-.1667em\lower.7ex\hbox{E}\kern-.125emX}}
\begin{document}

\title{Diffusion-based Speech Enhancement with Schr\"odinger Bridge and Symmetric Noise Schedule
}

\author{\IEEEauthorblockN{Siyi Wang}
\IEEEauthorblockA{
\textit{Logitech, EPFL}\\
Lausanne, CH}
\and
\IEEEauthorblockN{Siyi Liu}
\IEEEauthorblockA{
\textit{Logitech, EPFL}\\
Lausanne, CH}
\and
\IEEEauthorblockN{Andrew Harper}
\IEEEauthorblockA{
\textit{Logitech Europe}\\
Lausanne, CH}
\and
\IEEEauthorblockN{Paul Kendrick}
\IEEEauthorblockA{
\textit{Logitech Europe}\\
Lausanne, CH}
\and
\IEEEauthorblockN{Mathieu Salzmann}
\IEEEauthorblockA{
\textit{EPFL SDSC}\\
Lausanne, CH}
\and
\IEEEauthorblockN{Milos Cernak}
\IEEEauthorblockA{
\textit{Logitech Europe}\\
Lausanne, CH}
}

\maketitle

\begin{abstract}
Recently, diffusion-based generative models have demonstrated remarkable performance in speech enhancement tasks. However, these methods still encounter challenges, including the lack of structural information and poor performance in low Signal-to-Noise Ratio (SNR) scenarios. To overcome these challenges, we propose the \textbf{S}chr\"odinger \textbf{B}ridge-based \textbf{S}peech \textbf{E}nhancement (SBSE) method, which learns the diffusion processes directly between the noisy input and the clean distribution, unlike conventional diffusion-based speech enhancement systems that learn data to Gaussian distributions. To enhance performance in extremely noisy conditions, we introduce a two-stage system incorporating ratio mask information into the diffusion-based generative model. Our experimental results show that our proposed SBSE method outperforms all the baseline models and achieves state-of-the-art performance, especially in low SNR conditions. Importantly, only a few inference steps are required to achieve the best result.
\end{abstract}


\section{Introduction}




Current advancements have seen diffusion-based generative models achieving impressive outcomes in data generation tasks, extending their application to speech enhancement \cite{croitoru2023diffusion}. 
Initially introduced for image synthesis tasks, the denoising diffusion probabilistic model has demonstrated substantial capabilities in both generation and denoising, as noted in \cite{ho2020denoising}. 
The first application of a diffusion generative model to speech enhancement was proposed as the DiffuSE system~\cite{lu2021study} that enhances speech quality using a denoising diffusion probabilistic model (DDPM). To address a broader range of noises beyond Gaussian noise, improvements were made to DiffuSE, leading to the development of CDiffuSE ~\cite{lu2022conditional}. Other efforts employing score-based generative models, as outlined in \cite{richter2023speech} and \cite{serra2022universal}, have successfully produced higher-quality enhanced speech.

However, current diffusion-based generative methods suffer from the challenge of \textit{lacking structural information for inference}. Due to the inherent logic of the diffusion probabilistic model, the aforementioned diffusion models begin the inference process with Gaussian white noise or noisy speech mixed with strong Gaussian noise, which contains minimal or no structural information about the clean data distribution. Furthermore, for the models that start inference from the mixture of noisy speech and Gaussian noise, such as CDiffuSE, controlling the ratio of noisy speech and Gaussian noise still needs to be explored. 
Another challenge of current diffusion-based generative methods is their \textit{poor performance in low signal-to-noise ratio (SNR) conditions}. The diffusion-based generative model demonstrates a good ability to produce clean, high-quality speech in most conditions. However, in highly noisy environments, particularly when the SNR is below 0, the enhanced speech quality significantly degrades, yielding poor intelligibility, necessitating further improvement. 

To address the drawbacks, \textit{we propose to use i) the Schrödinger Bridge and ii) its extension with a conventional mask prediction model}.  The Schrödinger Bridge (SB) problem \cite{schrodinger1932theorie, chen2021likelihood}, is to seek an optimal way to transform one probability distribution into another arbitrary distribution. 
Recently, SB has been adopted to image reconstruction task \cite{liu20232} and text-to-speech synthesis task \cite{chen2023schrodinger}. 
Inspired by the SB concept, we apply the SB approach to speech enhancement, initiating the generative process directly from the noisy input. Compared to traditional diffusion-based speech enhancement methods, SB maintains more structural information on the initial state of the generative process. Furthermore, it also eliminates the need to balance Gaussian noise against noisy speech, offering a more direct and efficient pathway to speech enhancement.



In this paper, we propose the \textbf{S}chrodinger \textbf{B}ridge-based \textbf{S}peech \textbf{E}nhancement (SBSE) method within the complex STFT domain, which enables the direct generation of clean data from noisy speech.
The SBSE is grounded in a score-based generative framework and navigates through the forward and reverse processes as defined by certain Stochastic Differential Equations (SDEs). The SBSE initiates the reverse process directly from noisy speech, aiming to learn the nonlinear diffusion process from noisy to clean speech. NVIDIA recently explored both Variance Exploding (VE) SDE and Variance Preserving (VP) SDE, with VE showing better results~\cite{jukic24_interspeech}. Our method also uses the VE SDE structure but with the key difference of setting a symmetric noise scheduling, where the diffusion shrinks at both boundaries.

Besides, we combine the SB concept with a two-stage approach inspired by StoRM~\cite{storm23} and \cite{wang2023cross}. While we also utilize predictive models to aid generative models, our approaches diverge. We condition the diffusion process by combining the mask from the predictive model with the original noisy input, unlike StoRM, which uses only the predictive model's output. We opt for the magnitude ratio mask over the binary mask to provide more information to the generative model.
Incorporating a ratio mask enhances the quality of generated speech, especially under low SNR conditions.

\section{Background}
\subsection{Score-based Generative Models}
Diffusion models involve two processes: a forward process that transforms the data distribution $x_{0}$ into a prior distribution $x_T$, such as Gaussian distribution, through a predefined perturbing kernel $q_t(x_t)$ in $T$ steps, and a reverse process $q_t(x_{t-1}|x_t)$ that undoes the forward process. The score-based generative model (SGM) \cite{song2020score, song2020improved} builds on a continuous-time framework, leveraging stochastic differential equations (SDEs) for its forward and reverse process, which are described as
\begin{subequations}\label{eq:1}
    \begin{gather}
    d \mathbf x_t = \mathbf f(\mathbf x_t, t) dt + g(t) d\mathbf w_t \text{,}\\
    d \mathbf x_t = \left[ \mathbf f(\mathbf x_t, t) - g^2(t) \nabla \log p_t(\mathbf x_t) \right] dt + g(t) \mathbf dw_t \text{,}
    \end{gather}
\end{subequations}
where $f(x_t, t)$ is a vector-valued drift term, $g(t)$ is the diffusion coefficient that controls the amount of Gaussian noise introduced at each time step, $w$ refers to a standard Wiener process, and $t \in ({0,...,T})$. The forward and reverse processes share the same marginal distribution.

To generate the enhanced data through the reverse process from $t=T$ to $t=0$, a time-dependent neural network $s_\theta(x_t,t)$ parameterized by $\theta$ is employed to estimate the score function $\nabla_{\mathbf{x}} \log p(\mathbf{x})$. The model $s_\theta(x_t,t)$ is trained by a denoising score-matching objective \cite{vincent2011connection,song2020score} defined as
\begin{equation}\label{eq:2}
    \mathbb{E}_{t} \left[ \lambda(t) \mathbb{E}_{x_0} \mathbb{E}_{q(x_t|x_0)} \left[ \Vert s_{\theta}(x_t, t) - \nabla \log p(x_t|x_0) \Vert_{2}^{2} \right] \right] \text{,}
\end{equation}
where $\lambda(t)$ is positive weighting function, and $p(x_t|x_0)$ denotes the conditional transition determined by forward SDE.

\subsection{Schrödinger Bridge Model}
\subsubsection{Schrödinger Bridge Problem}
The SB problem \cite{schrodinger1932theorie, chen2021likelihood, leonard2013survey, wang2021deep} aims to optimize the transformation between two probability distributions 
over a fixed time, under the dynamics of a stochastic process. SB can be represented using the forward-backward SDEs 
\begin{subequations}\label{eq:3}
    \begin{gather}
    d \mathbf {x_t} = [f( \mathbf x_t, t) + g^2(t) \nabla \log \Psi_t (\mathbf{x_t})]dt + g(t)d \mathbf {w_t}, \label{eq:3a}\\
    d \mathbf {x_t} = [f( \mathbf x_t, t) - g^2(t) \nabla \log \hat{\Psi}_t (\mathbf{x_t})]dt + g(t)d \mathbf{ \tilde{w}_t}\label{eq:3b},
    \end{gather}
\end{subequations}
where $x_0$ and $x_T$ are drawn from the boundary distributions $p_A(x)$ and $p_B(x)$, respectively, and $f$ and $g$ are the same as the score-SDE process of Eq. (\ref{eq:1}). The nonlinear drifts $\nabla \log \Psi_t (\mathbf{x_t})$ and $\nabla \log \hat{\Psi}_t (\mathbf{x_t})$ can be described by following coupled partial differential equations (PDEs)
\begin{subequations}
    \begin{gather}
        \left\{
        \begin{array}{ll}\label{eq:4a}
            \frac{\partial \Psi(x)}{\partial t} = -\nabla \Psi^\top f - \frac{1}{2} \beta \Delta \Psi \\
            \frac{\partial \hat{\Psi}(x)}{\partial t} = -\nabla \cdot (\hat{\Psi} f) + \frac{1}{2} \beta \Delta \hat{\Psi} \text{,}
        \end{array}
        \right.  \\
        \text{s.t. } \Psi_0(x)\hat{\Psi}_0(x)=p_A(x), \ \Psi_T(x)\hat{\Psi}_T(x)=p_B(x) \text{.}
    \end{gather}
\end{subequations}
These additional nonlinear drift terms enable SB to extend data transportation beyond Gaussian priors.
To overcome the scalability and applicability challenges of the SB problem \cite{de2021diffusion, chen2021likelihood}, Liu et al. \cite{liu20232} proposed Image-to-Image Schr\"odinger Bridge (I$^2$SB), a simulation-free framework that learns the nonlinear diffusion processes between two given distributions.

\subsubsection{Simulation-free Framework}\label{section-sb background train}
Given the paired data, Liu et al. \cite{liu20232} developed a simulation-free methodology based on the SGM framework to efficiently tackle the SB problem. By conceptualizing $\Psi_t(\mathbf{x})$ and $\hat{\Psi}_t(\mathbf{x})$ as density functions, the drift terms $\hat{\Psi}_t(\mathbf{x})$ and ${\Psi}_t(\mathbf{x})$ effectively become the score functions respectively associated with the following linear SDEs
\begin{subequations}\label{eq:5}
    \begin{gather}
            d \mathbf x_t = \mathbf f(\mathbf x_t, t) dt + g(t) d\mathbf w_t, \text{           } x_0 \sim  \hat\Psi_0(x),\label{eq:5a}\\
            d \mathbf x_t = \mathbf f(\mathbf x_t, t) dt + g(t) d\mathbf w_t, \text{           } x_T \sim  \Psi_T(x).\label{eq:5b}
    \end{gather}
\end{subequations}
By leveraging these linear SDEs, the methodologies from the SGM framework can be employed to learn the score functions. 
To address the intractability of the boundary conditions $\hat \Psi_0(x)$ and $\Psi_T(x)$ introduced in Eq. (\ref{eq:4a}), Liu et al. set $p_A(x)$ as the Dirac delta distribution centered at $a$, defining $p_A(\cdot) := \delta_a(\cdot)$, thereby eliminating one of the couplings.

Taking $\Psi_t(\mathbf{x}_t|\mathbf{x}_0)$ and $\hat{\Psi}_t(\mathbf{x}_t|\mathbf{x}_T)$ as solutions to the Fokker-Planck equations and conditioning on Nelson's duality \cite{nelson2020dynamical}, the posterior distribution can be articulated in an analytic form when provided with boundary pair data \cite{sarkka2019applied}. Specifically
\begin{subequations} \label{eq:6}
    \begin{gather}
        q(x_t | x_0, x_T) = \mathcal{N}(x_t; \mu_t(x_0, x_T), \Sigma_t), \\
    \mu_t = \frac{\bar\sigma_t^2}{\bar\sigma_t^2 + \sigma_t^2} x_0 + \frac{\sigma_t^2}{\bar\sigma_t^2 + \sigma_t^2} x_T,
    \Sigma_t = \frac{\sigma_t^2 \bar \sigma_t^2}{\bar\sigma_t^2 + \sigma_t^2} \cdot I,
    \end{gather}
\end{subequations}
where $\sigma_t^2 := \int_0^t \beta_\tau d\tau$ and $\bar\sigma_t^{2} := \int_t^1 \beta_\tau d\tau$ are analytic marginal variances. During training, given initial and terminal conditions $x_0 \sim p_A(x_0)$ and $x_T \sim p_B(x_t|x_0)$, we can directly sample $x_t$ at any time step t without solving the nonlinear diffusion. The sampling mechanism employs the Denoising Diffusion Probabilistic Model (DDPM) sampler and can be written as the recursive posterior
\begin{equation}\label{eq:7}
    q(X_n | X_0, X_N) = \int \prod_{k=n}^{N-1} p(X_k | X_0, X_{k+1}) \, dX_{k+1}.
\end{equation}

According to Eq. (\ref{eq:3b}), we need $\nabla \log \hat{\Psi}_t (\mathbf{x_t})$ to conduct the reverse process, which is also the score function of Eq. (\ref{eq:5a}). Similar to Eq. (\ref{eq:2}) utilized for SGM, a neural network $s_\theta(x_t, t)$ parameterized by $\theta$ is deployed to estimate the score function $\nabla \log \hat{\Psi}_t (\mathbf{x_t})$, which leads to the loss function
\begin{equation} \label{eq:8}
    L := \left\Vert s_\theta(x_t, t) - \frac{x_t - x_0}{\sigma_t} \right\Vert \text{.}
\end{equation}

\section{Method}
\begin{figure*}[h]
\centering
\includegraphics[width=0.98\textwidth]{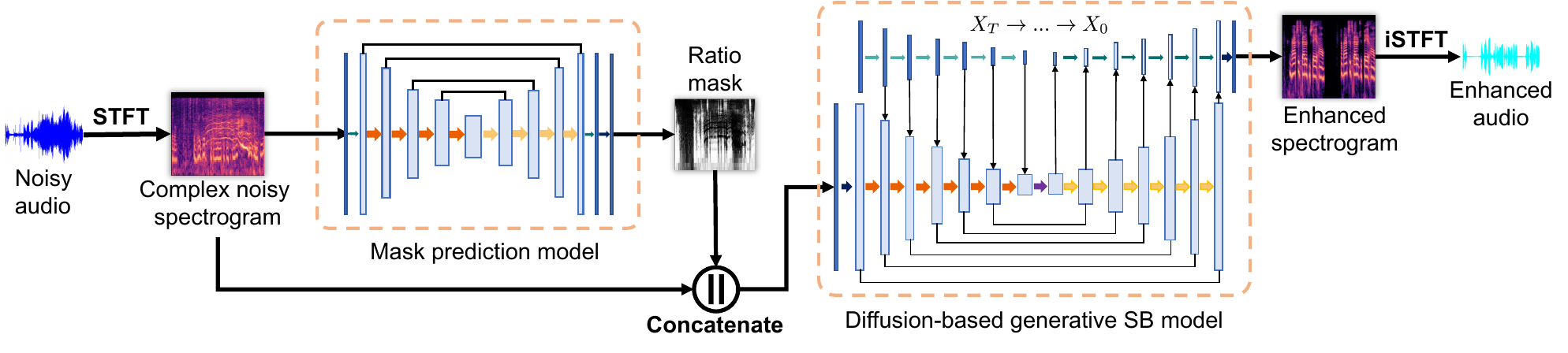}
\caption{Architecture of the proposed two-stage method. The Schrödinger Bridge (SB) can take a predicted mask as an auxiliary input.}
\label{fig:pipeline}
\end{figure*}

The two-stage method is shown in Fig.~\ref{fig:pipeline}.
\subsection{Ratio Mask Prediction Model}
The initial stage employs a ratio mask prediction U-Net model \cite{ronneberger2015u} that processes complex spectrograms to predict gain values. Oracle gains are defined as $g_{\text{Mag}} = |S|_\text{Mag}/|X|_\text{Mag}$~\cite{valin2020perceptually}, 
where $S$ signifies clean speech and $X$ represents noisy speech—a blend of clean speech and ambient noise.
The network contains 4 down- and 4 upsampling blocks, with a sigmoid activation. The predicted mask is produced as a single channel. The model parameters are optimized using the Mean Square Error (MSE) loss function between the oracle and estimated gains.


\subsection{Schr\"odinger Bridge Model}

\subsubsection{Training Task}
As outlined in Section \ref{section-sb background train}, the forward process of the SB can be conducted following Eq. \ref{eq:6}. The initial condition $x_0$ and terminal condition $x_T$ correspond to clean and noisy speech respectively. Training begins by establishing a noise schedule (${\beta_0, ...,\beta_T}$). We use a symmetric noise scheduling following suggestions from prior score-based models~\cite{de2021diffusion, chen2021likelihood,liu20232}, whereas~\cite{jukic24_interspeech} follows the noise scheduling of SB-TTS~\cite{chen2023schrodinger}. For each clean and noisy speech pair, a timestep $t$ is uniformly sampled from $({1,...,T})$. Subsequently, $x_t$ can be derived given a clean speech $x_0$ and a corresponding noisy speech $x_T$ following Eq. \ref{eq:6}. A neural network $s_\theta(x_t, t, M)$, which processes $x_t$, the time step $t$, and the ratio mask $M$, is then trained to optimize the loss function defined in Eq. \ref{eq:8}.

We utilize a network based on the U-Net structure \cite{ronneberger2015u,dhariwal2021diffusion} incorporating a progressive growing of the input that provides a downsampled input to every feature map within the U-Net's contracting path~\cite{viazovetskyi2020stylegan2}.
The network configuration for our study features six downsampling and six upsampling blocks, with channel sizes set as $128 \times [1,1,2,2,4,4]$. At each resolution level, two residual blocks derived from BigGAN \cite{brock2018large} are incorporated in the downsampling blocks and three in the upsampling blocks. The attention layers (\cite{vaswani2017attention}) are added at the resolution of $32 \times 32, 16 \times 16,8 \times 8$.

\subsubsection{Inference Procedure}
We utilize the DDPM sampler to sample the clean speech, as expressed in Eq. (\ref{eq:7}). \cite{liu20232} proves that the marginal density of the SB forward processes $q(x_t|x_0, x_1)$ is the marginal density of DDPM posterior $p(x_n|x_0, x_{n+1})$, thus the DDPM sampler can be effectively utilized to execute the reverse process of SB. When $f:=0$, $p(x_n|x_0, x_{n+1})$ has an analytic Gaussian form
\begin{equation}\label{eq:11}
    \mathcal{N}\left(x_n; \frac{\alpha_n^2}{\alpha_n^2 + \sigma_n^2} x_0 + \frac{\sigma_n^2}{\alpha_n^2 + \sigma_n^2} x_{n+1}, \frac{\sigma_n^2 \alpha_n^2}{\alpha_n^2 + \sigma_n^2} \cdot I \right)
\end{equation}
where $\alpha_n^2:= \int_{t_n}^{t_{n+1}} \beta_\tau = \sigma_{n+1}^2 - \sigma_n^2
$ is the accumulated variance between two consecutive time steps $(t_n, t_{n+1})$. The reverse process initiates from the noisy speech distribution, starting at $n=T$ with $x_n=x_T$.  With the accurate prediction of network $s_\theta(x_n, n, M)$, the $x_0$ can be reconstructed as $x_0 = x_n - \sigma_n s_\theta(x_n, n, M)$ (Eq. \ref{eq:8}). Leveraging the DDPM sampler as in Eq. \ref{eq:11}, we can infer $x_{n-1}$. The clean data $x_0$ can be iteratively sampled over all reverse steps. 

\section{Experiments}
\subsection{Experimental Setup}

\begin{table*}
\centering
\caption{Speech enhancement results obtained for the 2023 DNS challenge dataset. The values indicate means and 95\% confidence intervals. We mark the best results in \textbf{bold}; the second-best are \underline{underlined}.}
\resizebox{\textwidth}{!}{
    \begin{tabular}{l | ccc | ccc | ccc}
    \toprule
    \multirow{2}{*}{Method}  & \multicolumn{3}{c}{PESQ ($\uparrow$)} & \multicolumn{3}{c}{SI-SDR[dB] ($\uparrow$)} & \multicolumn{3}{c}{DNSMOS ($\uparrow$)} \\
    & $\text{SNR}=-5$ & $\text{SNR}=0$ & $\text{SNR}=[5,30]$ & $\text{SNR}=-5$ & $\text{SNR}=0$ & $\text{SNR}=[5,30]$ & $\text{SNR}=-5$ & $\text{SNR}=0$ & $\text{SNR}=[5,30]$ \\
    \midrule
    MetricGAN+  & $1.32_{\pm0.06}$ & $1.50_{\pm0.07}$ & $2.42_{\pm0.06}$ & $-6.47_{\pm0.97}$ & $-2.35_{\pm0.71}$ & $4.07_{\pm0.28}$ & $2.66_{\pm0.07}$ & $2.82_{\pm0.08}$ & $3.46_{\pm0.03}$ \\
    DeepFilterNet  & $1.40_{\pm0.06}$ & $1.60_{\pm0.08}$ & $2.74_{\pm0.06}$ & $5.99_{\pm0.82}$ & $9.03_{\pm0.69}$ & $17.47_{\pm0.43}$ & $3.27_{\pm0.08}$ & $3.51_{\pm0.07}$ & $3.85_{\pm0.02}$ \\
    CDiffuSE  &  $1.12_{\pm0.03}$ & $1.19_{\pm0.03}$ & $2.16_{\pm0.06}$ & $-3.88_{\pm0.86}$ & $2.15_{\pm0.67}$ & $10.12_{\pm0.22}$ & $2.57_{\pm0.05}$ & $2.74_{\pm0.06}$ & $3.27_{\pm0.03}$ \\
    SGMSE+ & $1.29_{\pm0.07}$ & $1.61_{\pm0.12}$ & $\mathbf{3.13_{\pm0.06}}$ & $0.39_{\pm1.25}$ & $7.21_{\pm1.22}$ & $22.14_{\pm0.55}$ & $3.17_{\pm0.10}$ & $3.46_{\pm0.09}$ & $3.85_{\pm0.03}$ \\
    StoRM & $\underline{1.43_{\pm0.08}} $ & $\underline{1.64_{\pm0.11}}$ & ${2.61_{\pm0.07}}$ & $4.45_{\pm 0.19}$ & $8.84_{\pm1.10}$ & $21.50_{\pm0.61}$ & $3.32_{\pm0.06}$ & $3.42_{\pm0.06}$ & $3.66_{\pm0.02}$ \\
    \midrule
    SBSE  & $1.42_{\pm0.09}$ & ${1.63_{\pm0.11}}$ & $2.86_{\pm0.07}$ & $\underline{7.88_{\pm0.90}}$ & $\underline{11.75_{\pm0.80}}$ & $\underline{22.89_{\pm0.52}}$ & $\underline{3.78_{\pm0.06}}$ & $\underline{3.85_{\pm0.05}}$ & $\mathbf{3.93_{\pm0.02}}$ \\
    SBSE-M  &$\mathbf{1.45_{\pm0.09}}$ & $\mathbf{1.69_{\pm0.11}}$ & $\underline{2.93_{\pm0.06}}$ & $\mathbf{8.31_{\pm0.88}}$ & $\mathbf{11.91_{\pm0.73}}$ & $\mathbf{22.99_{\pm0.53}}$ & $\mathbf{3.85_{\pm0.05}}$ & $\mathbf{3.91_{\pm0.05}}$ & $\mathbf{3.93_{\pm0.02}}$ \\
    
    \bottomrule
    \end{tabular}
}

\label{table-metrics}
\end{table*}

For model training, we utilize data instances from the 2023 Deep Noise Suppression (DNS) Challenge dataset \cite{dubey2023icassp}. These instances are created by randomly mixing speech and noise instances at SNR levels uniformly distributed between $[-5, 20]$ dB, with a sampling rate of 16 kHz. The training dataset consists of 60,000 audio instances, each 10 seconds long, totaling around 167 hours. For evaluation,  we prepared 100 independent speech-noise mixtures for each test dataset, covering 8 SNR levels from -5 dB to 30 dB in 5 dB steps.

For the data preprocessing, a window size of 32 ms, a hop length of 8 ms, and the Hann window are used to transform the waveform into a complex spectrogram. We randomly select the segment that lasts 256 frames from the complex spectrogram at each training step. Following previous work \cite{richter2023speech}, we apply the same amplitude transformation technique on the complex spectrogram to bring out the frequency bins with low energy, thereby balancing the data.

For mask prediction, the model was trained on two NVIDIA GeForce RTX 4070 Ti (12 GB memory each) for 70 epochs using the Adam \cite{kingma2015adam} optimizer with a learning rate of $10^{-4}$ and a batch size of 16. The SB model is trained on four NVIDIA A10G (24 GB memory each) for 100 epochs. We use the Adam optimizer with a $10^{-4}$ learning rate and batch size of $4 \times 6 =24$. We use the symmetric scheduling of $\beta$ adopted in \cite{liu20232, de2021diffusion} for model training. We set the number of inference steps to five, as this configuration has exhibited favorable results in both intrusive and non-intrusive metrics.

The performance of the baselines and our proposed speech enhancement methods is evaluated by \textbf{PESQ}~\cite{rix2001perceptual}, \textbf{SI-SDR}~\cite{le2019sdr}, \textbf{DNSMOS}~\cite{reddy2021dnsmos}, and \textbf{MUSHRA listening test}~\cite{schoeffler2018webmushra}. All metrics improve as their value increases.

\subsection{Baselines}

In evaluating the proposed methods, the~\textbf{SBSE} and its ratio mask extension~\textbf{SBSE-M} models, we compare them with two discriminative models \textbf{DeepFilterNetV3}~\cite{schroeter2023deepfilternet3} and~\textbf{MetricGAN+}~\cite{fu2021metricgan+}, and three diffusion-based models, \textbf{CDiffuSE}~\cite{lu2022conditional}, \textbf{SGMSE+}~\cite{richter2023speech} and two-stage model \textbf{StoRM}~\cite{storm23}.

We did not include NVIDIA SB-based baseline~\cite{jukic24_interspeech} as it was published shortly before this submission and without pre-trained models.

\subsection{Speech Quality Assessment}

\begin{figure}[h]
\centering
\includegraphics[width=0.48\textwidth]{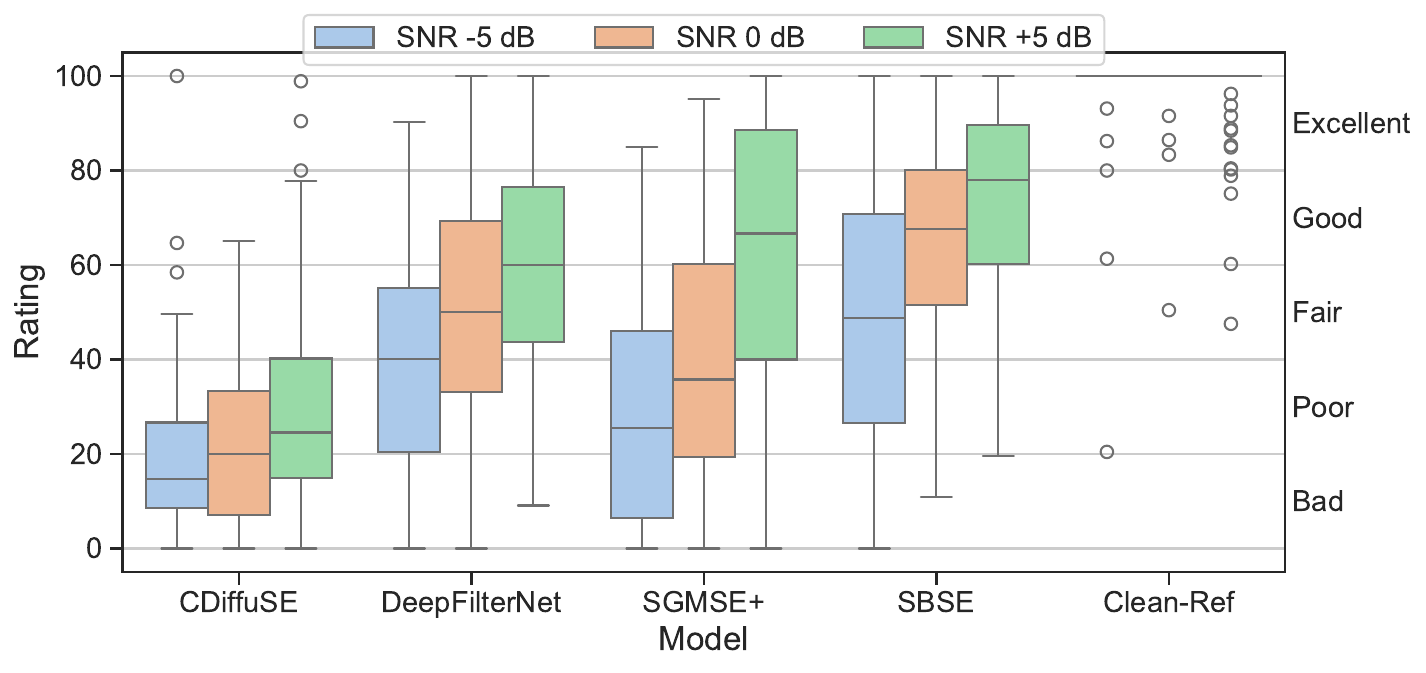}
\caption{MUSHRA subjective evaluation with 19 participants.}
\label{fig:mushra}
\end{figure}

Tab. \ref{table-metrics} presents the objective evaluation results on the DNS test set, specifically targeting low SNR conditions (SNR $\leq 0$). Fig.~\ref{fig:mushra} reports the outcomes of the MUSHRA listening test. Audio examples are available at\footnote{\url{glistening-lebkuchen-8e3c56.netlify.app}}. Based on these results, the following observations can be made:
\begin{itemize}
    \item {Compared with StoRM and SGMSE+ diffusion-based models, the SBSE-M outperforms them in low SNR environments while achieving comparable results in high SNR scenarios. Qualitative assessments indicate that SGMSE+ produces vocalizing artifacts, such as sounds in highly noisy scenarios resembling breathing and sighing. Unlike baselines, our approach rarely produces strong, pronounced, distorted artificial noises.}
    \item {Proposed SBSE-M and SBSE models outperform the discriminative approaches across all scenarios. Notably, in low SNR conditions, our methods exceed the DeepFilterNetV2, which ranks highest among the baseline models. In high SNR environments, our models excel, producing high-quality enhanced speech and demonstrating substantial improvements over discriminative methods.} 
    \item {As shown in Fig. \ref{fig:mushra}, the proposed SBSE system received the highest scores in the listening test under low SNR situations. 
    Especially in the most challenging condition ($SNR=-5$), SBSE produced fair-quality speech, while the other methods received poor scores. These results align with those presented in Table \ref{table-metrics}}.
    \item {In most scenarios, incorporating a ratio mask improves the quality of generated speech. The mask input boosts speech quality in low SNR situations by providing extra information to the network, thereby mitigating the effects of strong noise and the lack of detail in the original input.}
\end{itemize}

\subsection{Inference Speed Evaluation}
We also evaluated the sampling speed of our proposed methods and baseline models of ten 10-second audio files measured on an NVIDIA GeForce RTX 4070 Ti. The Number of Function Evaluations (NFE) for baseline models are adopted from their original paper. For our methods, we have configured the NFE to 5 for SBSE and SBSE-M. This configuration has been determined to provide satisfactory outcomes in our experiments while being computationally efficient. We use the real-time factor (RTF) to represent the inference speed, which indicates the ratio of the time to process the audio to the audio length. The fastest model was the discriminative DeepFilterNet with $0.026$ RTF. Our proposed SBSE and SBSE-M models were about 10 times slower with $0.21$ RTF. The generative baselines CDiffuSE and SGMSE+ achieved RTF $1.31$ and $2.11$, respectively. Compared to discriminative models, which require only one step for inference, our model trades off longer inference time for improved outcomes. Unlike diffusion-based models, SBSE operates with fewer steps, faster processing, and better qualitative results.

\section{Conclusions}
In this paper, we have revisited the Schr\"odinger Bridge-Based Speech Enhancement method and proposed the two-stage system integrating ratio mask information into the generative model. Our experiment results have shown that the SBSE model outperforms both discriminative and diffusion-based baseline models in low SNR conditions and not degrade signals in high SNR scenarios. Furthermore, we have demonstrated the significance of the ratio mask in enhancing speech quality under very noisy conditions. Additionally, our method is also faster compared to other diffusion-based models.

Although our proposed method achieves promising performance, it has limitations. The generative SB model occasionally produces phonetically accurate vocalizing sounds lacking linguistic meaning in extremely noisy regions; the current methods fail to restore the audio fully, which belongs to our future work.



\balance

\bibliographystyle{IEEEtran}
\bibliography{mybib}

\end{document}